\pdfoutput=1
\PassOptionsToPackage{dvipsnames}{xcolor}
\pdfoutput=1
\documentclass[sigconf, nonacm]{acmart}

\pdfoutput=1

\usepackage{natbib}
\usepackage{amsfonts} 
\usepackage{algorithm}
\usepackage{algpseudocode}
\usepackage{enumitem}

\usepackage{multirow}
\newcommand{\highest}[1]{\textcolor{Maroon}{{#1}}}%

\newcommand{\bs}{\mathbf{s}}
\newcommand{\ba}{\mathbf{a}}
\newcommand{\bu}{\mathbf{u}}

\newcommand{\bA}{\mathbf{A}}

\newcommand{\reb}{\mbox{\scriptsize reb}}

\AtBeginDocument{%
  \providecommand\BibTeX{{%
    \normalfont B\kern-0.5em{\scshape i\kern-0.25em b}\kern-0.8em\TeX}}}

\setcopyright{iw3c2w3g}
\copyrightyear{2022}
\acmYear{2022}
\acmDOI{10.1145/1122445.1122456}

\acmConference[SIGKDD '22]{SIGKDD '22: Conference on Knowledge Discovery and Data Science}{August 14--18, 2022}{Washington D.C., U.S.}
\acmBooktitle{SIGKDD '22: Conference on Knowledge Discovery and Data Science,
  August 14--18, 2022, Washington D.C., U.S.}
\acmPrice{XXXX}
\acmISBN{XXXX}

\begin{document}

\title{Graph Meta-Reinforcement Learning for Transferable Autonomous Mobility-on-Demand}

\author{Daniele Gammelli}
\affiliation{%
  \institution{Technical University of Denmark}
  \city{Kgs. Lyngby}
  \country{Denmark}
  }
\email{daga@dtu.dk}

\author{Kaidi Yang}
\affiliation{%
  \institution{Stanford University}
  \city{Stanford}
  \country{California}
  }
\email{kaidi.yang@stanford.edu}

\author{James Harrison}
\affiliation{%
  \institution{Google Research, Brain Team}
  \city{San Francisco}
  \country{California}
}
\email{jamesharrison@google.com}

\author{Filipe Rodrigues}
\affiliation{%
  \institution{Technical University of Denmark}
  \city{Kgs. Lyngby}
  \country{Denmark}
  }
\email{rodr@dtu.dk}

\author{Francisco C. Pereira}
\affiliation{%
  \institution{Technical University of Denmark}
  \city{Kgs. Lyngby}
  \country{Denmark}
  }
\email{camara@dtu.dk}

\author{Marco Pavone}
\affiliation{%
  \institution{Stanford University}
  \city{Stanford}
  \country{California}
  }
\email{pavone@stanford.edu}


\begin{abstract}
    Autonomous Mobility-on-Demand (AMoD) systems represent an attractive alternative to existing transportation paradigms, currently challenged by urbanization and increasing travel needs.
    By centrally controlling a fleet of self-driving vehicles, these systems provide mobility service to customers and are currently starting to be deployed in a number of cities around the world. 
    Current learning-based approaches for controlling AMoD systems are limited to the \emph{single-city} scenario, whereby the service operator is allowed to take an unlimited amount of operational decisions within the same transportation system.
    However, real-world system operators can hardly afford to fully re-train AMoD controllers for every city they operate in, as this could result in a high number of poor-quality decisions during training, making the single-city strategy a potentially impractical solution.
    To address these limitations, we propose to formalize the multi-city AMoD problem through the lens of meta-reinforcement learning (meta-RL) and devise an actor-critic algorithm based on recurrent graph neural networks.
    In our approach, AMoD controllers are explicitly trained such that a small amount of experience within a new city will produce good system performance.
    Empirically, we show how control policies learned through meta-RL are able to achieve near-optimal performance on unseen cities by learning rapidly adaptable policies, thus making them more robust not only to novel environments, but also to distribution shifts common in real-world operations, such as special events, unexpected congestion, and dynamic pricing schemes.
\end{abstract}




\maketitle

   
\section{Introduction}

More than half of the world's population now lives in urban areas, and this proportion is expected to reach $68\%$ by 2050 \cite{UNDepEconAff2021}.
This growth will result in an increase in urban travel, and as such, the externalities that this travel produces will become more severe: production of greenhouse gases, dependency on oil, and traffic congestion, to name a few.
Increasingly congested transportation systems are considered responsible for $30\%$ of emissions within urban environments and 50\% of health-related cost due to car-generated air pollution \cite{OECD2014}. Hence, current transportation paradigms based on private cars are widely regarded as an unsustainable solution for future personal mobility. 
Therefore, cities face the challenging task of devising transportation systems that can sustainably match the growing mobility needs and reduce environmental harm.
   
The concept of mobility-on-demand (MoD) has the potential to provide the convenience of individual-scale point-to-point transportation while reducing the severe negative externalities of private vehicles.
Specifically, customers in an MoD system typically request a one-way ride from an origin to a destination and are served by a fleet of shared vehicles. 
In this way, due to higher vehicle utilization, MoD systems have the potential to address issues such as pollution and saturated parking spaces while satisfying the needs of personal mobility \cite{ChanShaheen2012}.
However, one of the major operational challenges of MoD systems lies in their tendency to become unbalanced due to asymmetric transportation demand (e.g., commuting into downtown in the morning and vice-versa in the evening). 
Suboptimal rebalancing can jeopardise the benefits of MoD, since it can increase passenger waiting time or cause traffic congestion by creating unnecessary rebalancing trips with empty vehicles \cite{OhLentzakisEtAl2021}.  
To address this problem, technological advancements in the field of autonomous driving offer a potential solution: autonomous mobility-on-demand (AMoD).
Specifically, the use of AVs within MoD systems enables service operators to centrally coordinate vehicles, thus simultaneously assigning passengers to AVs and rebalancing the fleet by relocating customer-free AVs.
Driven by such promising benefits, various companies started to develop the technology, and some are already providing autonomous ride-hailing services. 
However, controlling AMoD systems potentially entails the routing of large robotics fleets within complex transportation systems, thus making the AMoD control problem an open challenge.  

Among other approaches, methods based on reinforcement learning (RL) cast the AMoD control problem as the problem of learning to control a dynamical system from experience.
A common assumption underlying the majority of RL methods for AMoD systems is that the learned control policies will be operating in a single city.
On the contrary, real-world deployment of AMoD systems will likely be similar to current ride-hailing services, wherein a limited number of providers operate in thousands of cities worldwide.
Crucially, the single-city scenario would result in system operators having to fully train an AMoD control policy every time it is deployed to a new city, or more broadly, every time there is a substantial change in the transportation system.
This approach would be impractical in a real-world setup, where taking sub-optimal decisions is expensive, and sample efficiency is paramount.
   
In this work, we propose to formalize the \emph{multi-city} AMoD control problem from a meta-RL perspective, with the goal of enabling agents to leverage previous experience in order to quickly adapt to new cities. 
In particular, we define each city as representing a new task for the AMoD controller (i.e., the agent), and propose meta-RL as a mathematical formalism to learn policies capable of producing good performance on new cities (i.e., environments) with only a limited amount of online operational decisions within a given transportation network (i.e., interactions).
We argue that control policies learned through meta-RL exhibit a number of desirable properties, and propose recurrent graph neural networks as a general approach to learn a single AMoD control policy capable of leveraging previous experience and adapt to novel geographies and conditions.
Notably, compared to other architectures, graph-based computations pair naturally with multi-city AMoD systems, through their ability to process transportation networks of varying size and connectivity (e.g., MLPs require inputs with fixed dimensionality), while avoiding non-trivial aggregations (e.g. CNNs require pixel-grid representations).
It is worth noting that although this paper focuses on AMoD systems, the proposed approach can be applied to existing MoD systems (without AVs) if there are compliant drivers, e.g., drivers paid to strictly follow the instructions given by the operator \cite{YangTsaoEtAl2021}. 

\noindent \textbf{Related work.} Within existing literature, AMoD systems can be coordinated with simple rule-based heuristics \cite{HylandMahmassani2018,LevinKockelmanEtAl2017}, model predictive control (MPC) approaches ~\cite{ZhangPavone2016,IglesiasRossiEtAl2018}, and RL approaches~\cite{FluriRuchEtAl2019,HollerVuorioEtAl2019,MaoLiuEtAl2020,FengGluzmanEtAl2020,SkordilisHouEtAl2021,GammelliEtAl2021}. Interested readers can refer to \cite{ZardiniLanzettiEtAl2021} for a comprehensive survey of the analysis and control of AMoD systems and \cite{QinZhuEtAl2021} for RL approaches for ridesharing systems. 
For brevity, we only review RL approaches to rebalance vehicles in a centralized manner.

A key challenge with centralized approaches to the AMoD control problem is the high-dimensionality of the joint action space for all vehicles. 
To address this scalability challenge, Mao et al. \cite{MaoLiuEtAl2020} developed an actor-critic algorithm to determine an aggregated number of vehicles traveling between each pair of zones, thus obtaining an action space independent of the number of vehicles. 
However, this work does not consider specific neural network architectures that can exploit the graph structure of road networks, nor address the challenge when the number of zones is large. 
Fluri et al. \cite{FluriRuchEtAl2019} devised a cascaded Q-learning approach that represents the transportation network in a tree-like cascaded structure and derives an RL agent for each node. This approach can significantly reduce the number of state-action pairs for each agent, allowing more efficient learning. 
However, the learned agents can only take reactive decisions based on the current vehicle distribution, hence may not perform well in dynamic scenarios with time-varying travel demand and traffic conditions. 
Gammelli et al. \cite{GammelliEtAl2021} developed a deep RL approach based on graph neural networks to exploit the connectivity embedded in the transportation network to learn a policy that is scalable and near-optimal. Overall, the aforementioned RL approaches can ensure computational tractability without significantly compromising optimality.

However, to the best of our knowledge, existing RL approaches are limited to the single-city scenario, wherein an RL agent is trained in a single transportation network or geographical region. 
Although transferability across cities is discussed through experiments in several works (e.g., \cite{GammelliEtAl2021,SkordilisHouEtAl2021}), there lacks a systematic and formalized methodology that can explicitly train controllers capable of adapting to new cities. 
Several works employ meta-learning to develop prediction models for traffic states \cite{ZhangLiEtAl2022} and travel demand \cite{YaoLiuEtAl2019} that are adaptable to multiple cities, however, these works solely develop predictive models without considering the adaptation of the control strategy, and it is not clear how the insights can be applied to coordination algorithms of AMoD fleets.   

\noindent \textbf{Paper contributions.} The contributions of this paper are threefold.
First, we formalize the AMoD control problem through the lens of meta-RL and propose this as a viable solution to enable deep RL agents to operate within multi-city AMoD systems.
By doing so, the agent can exploit previous experience (e.g., acquired from cities in which the system is already operating) in order to adapt to novel environments through small amount of experience.

Second, we define a novel neural architecture for a deep meta-RL agent whose policy (and critic) leverage (i) their recurrent structure for effective meta-learning and temporal adaptation, and (ii) the relational representative power of graph neural networks to process transportation networks of varying size and connectivity. 

Third, this works highlights how policies learned through meta-RL exhibit a series of desireable properties critical for real-world deployment of AMoD systems.
In particular, through an empirical study based on real-world trip data from different cities around the world, we show how the proposed framework achieves close-to-optimal performance when deployed in unseen cities (i.e., characterized by completely different geographies, demand patterns, travel times, etc.), with minimal interaction.
Specifically, results show the emergence of \emph{generally adaptive} behavior in the presence of changes in distribution, making the resulting policy more robust towards common disturbances such as (i) special events (i.e., sudden peaks in demand), (ii) congestion (i.e., sudden increase in travel times), and (iii) changes in pricing scheme.

\section{BACKGROUND}
\label{sec:background}
In this section, we introduce notation and theoretical foundations underlying our work in the context of meta-RL (Section \ref{subsec:meta_rl}) and graph neural networks (Section \ref{subsec:graph_neural_networks}).

\subsection{The Meta-Reinforcement Learning Problem}
\label{subsec:meta_rl}
RL aims to solve the problem of learning to control a dynamical system from experience \cite{SuttonBarto1998}.
Formally, we will refer to a dynamical system as being entirely determined by a (fully-observed) Markov decision process (MDP) $\mathcal{M} = (\mathcal{S}, \mathcal{A}, P, \rho, r, \gamma)$, where $\mathcal{S}$ is a set of all possible states $\mathbf{s} \in \mathcal{S}$, which may be either discrete or continuous, $\mathcal{A}$ is the set of possible actions $\mathbf{a} \in \mathcal{A}$, also discrete or continuous, $P$ describes the dynamics of the system through a conditional probability distribution of the form $P(\bs_{t+1} | \bs_{t}, \ba_{t})$, $\rho$ represents the initial state distribution $\rho(\bs_0)$, $r : \mathcal{S} \times \mathcal{A} \rightarrow \mathbb{R}$ defines the reward function and $\gamma \in (0, 1]$ is a scalar discount factor.
At a high level, the goal of a decision-making agent (such as the component of our framework doing rebalancing) is to learn a policy $\pi(\ba_t | \bs_t)$ (i.e., a probability distribution over actions given a state) by interacting with the MDP $\mathcal{M}$ (i.e., the environment) under the objective of maximizing the expected sum of cumulative rewards $\mathbb{E}_{\bs_t, \ba_t \sim \pi} \left[\sum_t \gamma^t r(\bs_t, \ba_t) \right]$. 

In its most general formulation, meta-learning addresses the problem of learning to learn, or in other words, automatically discovering learning algorithms that are more efficient and effective than learning tabula rasa.
The key idea is to define algorithms able to leverage data from previous tasks (i.e., meta-training) in order to acquire a learning procedure that can quickly adapt to new tasks (i.e., meta-test).
The common assumption underlying all meta-learning algorithms is that tasks are drawn from the same task distribution $p(\mathcal{T})$ and share a common structure that can be exploited for fast learning.
In our work, we will refer to a task as an MDP $\mathcal{T}_k = (\mathcal{S}_k, \mathcal{A}_k, P_k, \rho_k, r_k, \gamma_k)$.
Given a small dataset of task-specific experience $\mathcal{D}_{\mathcal{T}_k}$ (e.g. a small number of interactions with a fixed MDP), the goal of an agent is to learn a policy $\pi(\ba_t | \bs_t, \mathcal{D}_{\mathcal{T}_k})$ by optimizing for expected performance over $p(\mathcal{T})$, $\mathbb{E}_{\mathcal{T}_k \sim p(\mathcal{T}), \bs_t, \ba_t \sim \pi} \left[\sum_t \gamma^t r(\bs_t, \ba_t) \right]$, such that it can generalize to novel, unseen tasks.

At a high level, meta-RL strategies can be divided into two families of methods: gradient-based and black-box.
Among gradient-based approaches, model-agnostic meta-learning (MAML) \cite{FinnAbbeelEtAl2017} aims to learn the initial configuration of policy parameters (typically a neural network) such that taking a small number of gradient descent steps leads to few-shot generalization to new tasks.
Thus, this family of methods leverages the inductive bias of gradient-based optimization to facilitate adaptation to a new task.
On the other hand, the learning rule in black-box methods is considered a learnable module and is learned together with all other model parameters.
A typical approach is to parametrize the learning rule through a recurrent neural network \cite{hochreiterYounger2001, WangNelsonEtAl2016, DuanEtAl2016}, such that the agent is encouraged to sequentially integrate all the information it has received through $\mathcal{D}_{\mathcal{T}_k}$ and adapt its strategy continually.

An approach of particular interest for this work is RL$^2$ \cite{DuanEtAl2016}.
Let $n$ denote the total number of episodes of interaction with a specific task $\mathcal{T}_k$, and the term \emph{trial} a series of episodes within a fixed task.
The process of interaction between an agent and the environment is as follows: for each trial, we draw a task $\mathcal{T}_k \sim p(\mathcal{T})$ and for each of the $n$ episodes we draw an initial state $\bs_0 \sim \rho_0(\bs_0)$.
At every timestep $t$, the current state $\bs_t$, previous action $\ba_{t-1}$, reward $r_{t-1}$ and termination flag $d_{t-1}$ (e.g., we assume $d_t=1$ if the episode has terminated and 0 otherwise) are concatenated to form the input to the policy, which, parametrized as a recurrent model, is conditioned on the current hidden state $h_t$ to generate the action $\ba_t$ and next hidden state $h_{t+1}$.
Under this formulation, the goal of an agent is to maximize the expected sum of rewards over an entire trial rather than a single episode.
A key feature of this framework is the fact that the hidden state of the policy is preserved across episodes and only reset between trials.
In this way, the agent is encouraged to develop exploration strategies across consecutive episodes in order to quickly adapt to a new task.

\subsection{Graph Neural Networks}
\label{subsec:graph_neural_networks}
Let us a define a directed graph $\mathcal{G} = (\mathcal{V}, \mathcal{E}, \bu)$, where $\mathcal{V} = \{v_i\}_{i=1:N_v}$, $\mathcal{E} = \{e_k\}_{k=1:N_e}$ represent the set of nodes and edges of $\mathcal{G}$ (each described by a vector of features), with $\bu$ being a vector of global features.
Graph networks (GN) \cite{BattagliaPascanuEtAl2016}, define a class of functions for relational reasoning over graph-structured data.
The GN framework is typically characterized by units of "graph-to-graph" operations, or GN \emph{blocks}, which perform computations over the structure of an input graph and return a graph as output.
At its core, a GN block contains three sub-modules: edge-wise ($f_e$), node-wise ($f_v$) and global ($f_u$) update functions.
Concretely, when a graph is provided as input to a GN block, the computations proceed from the edge, to the node, to the global level, returning a graph with the same structure but updated edge, node and global features.
An appealing property of the GN framework is its flexibility and compositional structure, both in terms of what information is made available as input to its functions, as well as how edge, node and global updates are produced (e.g., to deal with sequential graph-structured data, we could define update functions based on RNNs).

\section{Control of Single-City AMoD Systems}
\label{sec:control_of_single_city_amod_systems}
In this section, we introduce the notation characterizing both the transportation network and AMoD controller  and define a three-step framework to control arbitrary AMoD systems.

\noindent \textbf{Problem Definition}. 
An AMoD operator coordinates $M$ taxi-like fully autonomous vehicles to provide on-demand mobility services on a transportation network represented by a graph $\mathcal{G} = (\mathcal{V}, \mathcal{E})$, where $\mathcal{V}$ represents the set of stations (e.g., pick-up or drop-off locations) and $\mathcal{E}$ represents the set of links connecting two adjacent stations. Let $N_v = |\mathcal{V|}$ denote the number of stations.
The time horizon is discretized into a set of discrete intervals $\mathcal{I}=\{1,2,\cdots, T\}$ of a given length $\Delta T$. 
AVs traveling between station $i \in \mathcal{V}$ and station $j\neq i \in \mathcal{V}$ starting from each time step $t$ are controlled by the operator to follow the shortest path, with a travel time\footnote{Travel times are assumed to be given and independent of the control of the AMoD fleet. This is a commonly adopted assumption that applies to transportation networks where the AMoD fleet constitutes a small proportion of the entire vehicle population  and imposes marginal impact on traffic congestion. } of $\tau_{ij}^t \in \mathbb{Z}_+$ time steps and a travel cost of $c_{ij}^t$ (e.g., as a function of travel time). 


Passengers make trip requests at each time step.
The requests with origin-destination (OD) pair $(i, j)\in \mathcal{V}\times \mathcal{V}$ submitted at time step $t \in \mathcal{I}$ are characterized\footnote{We assume passenger demand and prices to be independent of the coordination of AMoD fleets. We further assume that passengers not matched with any vehicles within a time step will leave the system. These assumptions can be relaxed in future work by training the proposed RL model in an environment incorporating demand modeling and surge pricing.} by demand $d_{ij}^t$ and price $p_{ij}^t$. 
With these trip requests, the operator dynamically matches passengers to vehicles, and the matched vehicles will deliver passengers to their destinations. Idle vehicles not matched with any passengers will be controlled by the operator to either stay at the same station or rebalance to other stations. 
We denote $x_{ij}^t \in \mathbb{N}$ as the passenger flow, i.e. the number of passengers traveling from station $i$ to station $j$ at time step $t$ that are successfully matched with a vehicle, and $y_{ij}^t \in \mathbb{N}$ as the rebalancing flow, i.e., the number of vehicles rebalancing from station $i$ to station $j$ at time step $t$.


\begin{figure*}[t]
      \centering
     \includegraphics[width=\textwidth]{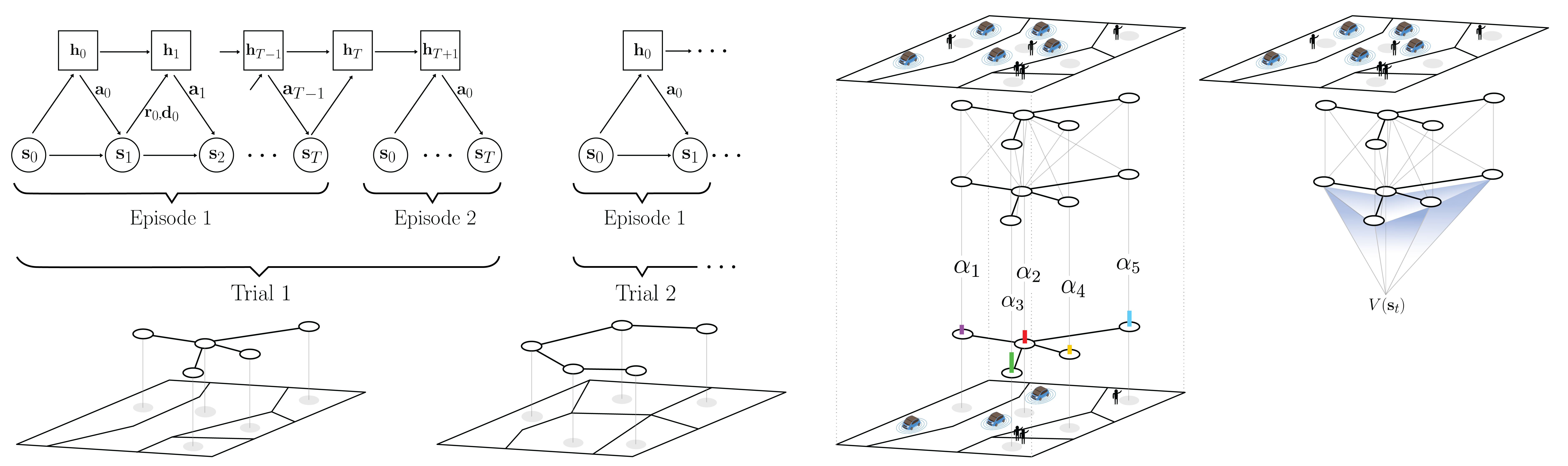}
      \caption{(Left) Interaction between the agent (squares) and the environment. Each trial is characterized by $n$ episodes within a fixed transportation network. (Right) Both actor and critic update raw graph representations of the transportation network to compute (i) a desired distribution of idle vehicles and (ii) an estimate of the value function, respectively.}
      \label{fig:3-step-framework}
   \end{figure*}
   
\noindent \textbf{A Three-Step Framework}.
As in \cite{GammelliEtAl2021, FluriRuchEtAl2019}, we formulate the AMoD control problem as a three-step decision-making framework, whereby the control of an AMoD fleet follows three steps: matching, RL for rebalancing, and post-processing. 
This three-step framework, as shown in the rest of this section, has the advantage of reducing the action space from $N_v^2$ to $N_v$, since 
the learned policy defines an action at each node as opposed to along each OD pair (as in the majority of literature). 
In what follows, we provide a detailed description for the framework. 

First, in the the matching step, the operator assigns vehicles to customers and obtains passenger flows $\{x_{ij}^t\}_{i,j\in\mathcal{V}}$ by solving the following  assignment problem:
\begin{subequations}
\begin{align}
    \max_{\{x_{ij}^t\}_{i,j\in\mathcal{V}}} \quad& \sum_{i,j\in\mathcal{V}}x_{ij}^t(p_{ij}^t- c_{ij}^t) \label{eq:matching_obj}\\
    \rm{s.t.} \,\,\,\,\, \quad& 0\leq x_{ij}^t \leq d_{ij}^t, ~i,j\in\mathcal{V}, \label{eq:matching_con1}\\
    & \sum_{j \in \mathcal{V}} x_{ij}^t \leq M_{i}^t, ~i\in\mathcal{V}, \label{eq:matching_con2}
\end{align}\label{eq:matching}
\end{subequations}
where the objective function (\ref{eq:matching_obj}) represents the profit of passenger assignment calculated as the difference between revenue and cost, the constraint (\ref{eq:matching_con1}) ensures that the passenger flow is non-negative and upper-bounded by the demand, the constraint (\ref{eq:matching_con2}) represents that the total passenger flow does not exceed the number of vacant vehicles $M_i^t$ at station $i$ at time step $t$. Notice that the constraint matrix of the assignment problem (\ref{eq:matching}) is totally unimodular\cite{Nemhauser1988}, hence the resulting passenger flows are integral as long as the demand is integral. 

Second, the RL step aims to determine the desired idle vehicle distribution $ \ba_{\reb}^t = \{a_{\reb, i}^t\}_{i \in \mathcal{V}}$, where $a_{\reb, i}^t \in [0,1]$ defines the percentage of currently idle vehicles to be rebalanced towards station $i$ in time step $t$, and $\sum_{i \in \mathcal{V}} a_{\reb, i}^t = 1$. 
With desired distribution $\ba_{\reb}^t$, denote $\hat{m}_i^t=\lfloor a_{\reb,i}^t\sum_{i\in\mathcal{V}}{m_i^t}\rfloor$ as the number of desired vehicles, where $m_i^t$ represents the actual number of idle vehicles after matching in region $i$ at time step $t$. Here, the floor function $\lfloor\cdot\rfloor$ is used to ensure that the desired number of vehicles is integral and always available ($\sum_{i\in\mathcal{V}}\hat{m}_i^t \leq \sum_{i\in\mathcal{V}}m_i^t$).

Third, the post-processing step converts the desired distribution into actionable rebalancing flows $\{y_{ij}^t\}_{i\neq j\in \mathcal{V}}$ by solving a minimal rebalancing-cost problem:
\begin{subequations}
\begin{align}
    \min_{\{y_{ij}^t\}_{i\neq j\in \mathcal{V}}\in \mathbb{N}^{|\mathcal{V}|\times (|\mathcal{V}|-1)}} \quad& \sum_{i\neq j\in \mathcal{V}}c_{ij}^t y_{ij}^t \label{eq:reb_obj}\\
    \rm{s.t.}  \quad \quad \quad \,
    & \sum_{j\neq i}(y_{ji}^t - y_{ij}^t) + m_i^t \geq \hat{m}_i^t , ~i\in\mathcal{V},\label{eq:reb_con1} \\
    \quad& \sum_{j\neq i} y_{ij}^t \leq m_i^t,~i\in\mathcal{V},\label{eq:reb_con2}
\end{align}\label{eq:reb}
\end{subequations}
where the objective function (\ref{eq:reb_obj}) represents the rebalancing cost, constraint (\ref{eq:reb_con1})  ensures that the resulting number of vehicles is close to the desired number of vehicles, and constraint (\ref{eq:reb_con2}) ensures that the total rebalancing flow from a region is upper-bounded by the number of idle vehicles in that region. 


\section{Meta-RL for AMoD}
\label{sec:meta_rl_for_amod}
In this section, we first formalize a black-box meta-RL method in the context of multi-city AMoD systems (Section \ref{subsec:rl2_for_amod}). 
We then formulate the AMoD rebalancing problem as an MDP (Section \ref{subsec:amod_rebalancing_mdp}) and introduce a recurrent graph network architecture for effective meta-learning over graphs (Section \ref{subsec:recurrent_gn}).

\subsection{RL$^2$ for Multi-City AMoD Systems}
\label{subsec:rl2_for_amod}
We now describe our formulation, which casts learning to control multi-city AMoD systems as a meta-RL problem. Our approach builds on RL$^2$ \cite{DuanEtAl2016}, a recurrence-based meta-learner. This was chosen over an optimization-based meta-learner (such as MAML \cite{FinnAbbeelEtAl2017}) for several reasons. First, recurrence-based models offer both flexibility of use---they are capable of recursively incorporating histories of different length---as well as simplicity of implementation. Second, recurrent models have preferable computational complexity, both in meta-training (due to avoiding higher-order derivatives) and during online operation. Finally, recurrent models have recently been shown to provide competitive performance on meta-learning problems and partially observed MDPs \cite{NiEysenbachEtAl2021}. 

We assume knowledge of a set of MDPs, or tasks, denoted by $\mathcal{T}$, and a distribution over them $p(\mathcal{T})$.
In our work, every task is characterized by a different city\footnote{While we focus on inter-city generalization, the notion of task can easily be extended to, e.g., different sub-city, regions, different times of the year, scenarios with changing demand and supply patterns, etc.} in which the AMoD controller needs to operate.
Each task is thus characterized by a unique transportation network (i.e., underlying graph topology, number of areas, travel times, etc.), demand pattern, pricing scheme and AMoD fleet size.
Despite the many differences, tasks share the same common structure: the way in which an AMoD fleet \emph{operates} in order to serve a given transportation demand, thus making it an appealing problem for meta-learning.

We use $n$ to denote the total number of episodes the agent is allowed to spend with a specific city.
Thus, the term \emph{trial} describes a series of $n$ such episodes.
The process of interaction between the agent and the environment is illustrated in Figure \ref{fig:3-step-framework}.
Under this formulation, the objective is to maximize the expected total reward accumulated during an entire trial, thus encouraging the agent to acquire a learning algorithm capable of quickly adapting to novel environments with limited amount of experience.

\subsection{AMoD Rebalancing MDP}
\label{subsec:amod_rebalancing_mdp}
As introduced in Section \ref{sec:control_of_single_city_amod_systems}, we attempt to learn a behavior policy to compute the desired distribution of idle vehicles through RL (Step 2).
Let us formulate the problem of rebalancing AMoD systems as an MDP $\mathcal{T}_{k}^{\reb} = (\mathcal{S}^{\reb}, \mathcal{A}_{k}^{\reb}, P_{k}^{\reb}, r^{\reb}, \gamma)$.

\emph{Action Space} ($\mathcal{A}_{\reb}$): We consider the problem of determining the \emph{desired idle vehicle distribution} $\ba_{\reb}^t$
, indicating the percentage of idle vehicles to be rebalanced from each station.

\emph{Reward} ($r_{\reb}$): We select the reward function in the MDP as the AMoD operator's profit within the current time step $t$, defined as the difference between the revenue of serving passengers and cost incurred by all vehicles with or without passengers, written as:  
\begin{equation}
    r_{\reb} = \sum_{i,j \in \mathcal{V}} x_{ij}^t (p_{ij}^t - c_{ij}^t) -  \sum_{i\neq j\in \mathcal{V}} y_{ij}^tc_{ij}^t,
\end{equation}

\emph{State Space} ($\mathcal{S}_{\reb}$): We define the state in the rebalancing MDP to contain the information needed to determine proactive rebalancing strategies, including (1) the structure of the transportation network characterized by its adjacency matrix $\bA$, and (2) station-level information by means of a feature matrix $\mathbf{X}$ characterized by:
\begin{itemize}[leftmargin=*]
    \item Current availability of idle vehicles in each station $m_i^t \in [0, M], \forall i \in \mathcal{V}$, as well as \emph{projected} availability of idle vehicles $\{m_i^{t^{'}}\}_{t^{'}=t, \ldots, t+H}$ estimated from previously assigned passenger and rebalancing trips, where $H$ represents the planning horizon.
    \item Observed transportation demand $\{d_{ij}^t\}_{i,j\in\mathcal{V}}$ and corresponding prices $\{p_{ij}^t\}_{i,j\in\mathcal{V}}$ and cost $\{c_{ij}^t\}_{i,j\in\mathcal{V}}$ at the current time step $t$. In this work, we take a fully model-free approach by assuming no knowledge of future transportation demand in the system, however, our approach could naturally be extended to process demand estimates from a predictive model, such as \cite{GammelliEtAl2020}.
    \item State $\bs_t$, action $\ba_{t-1}$, reward $r_{t-1}$, and termination flag $d_{t-1}$, aiming to address meta-learning through recurrence (see Figure \ref{fig:3-step-framework})\footnote{To ensure consistency in the input dimensions, we use placeholder values as initial input to the policy (i.e., a vector of zeros).}.
\end{itemize}
By means of this definition of the state space, we provide the behavior policy with meaningful information for it to capture statistics of the current and estimated future state of the MoD system, together with operational provider information and performance.

\smallskip\emph{Dynamics} ($P_{\reb}$):  The dynamics in the rebalancing MDP describe passenger arrival and evolution of state elements. First, passenger arrival for each OD pair is defined as an inhomogeneous Poisson process (i.e., with time-varying arrival rate estimated from real data) independent of the arrival processes of other OD pairs and the rebalancing action. Second, the evolution of state elements is defined as: (1) the estimated availability $\{m_i^{t^{'}}\}_{t^{'}=t, \ldots, t+T}$ is determined based on vehicle conservation, i.e., the number of idle vehicles at station $i$ at time step $t+1$, $m_i^{t+1}$, can be calculated as the number of idle vehicles at time $t$ ($m_i^{t}$) plus the number of incoming vehicles minus the number of outgoing vehicles (either for passenger delivery or rebalancing); (2) trip price $p_{ij}^t$ and cost $c_{ij}^t$ are assumed to be externally decided and known beforehand (hence, independent from the actions selected by the behavior policy).


\subsection{Recurrent GN for Rebalancing}
\label{subsec:recurrent_gn}
Having introduced the AMoD rebalancing MDP, this section introduces a temporal graph neural network architecture characterizing both the policy $\pi_{\theta}(\ba_t | \bs_t, \mathcal{D}_{\mathcal{T}_k})$ and value function estimator $V_{\phi}(\bs_t | \mathcal{D}_{\mathcal{T}_k})$ within the proposed Advantage Actor-Critic (A2C) \cite{MnihPuigdomenechEtAl2016} formulation.
Algorithm \ref{alg:gn} provides a schematic illustration.

In our work, we parametrize the actor and critic of A2C with two symmetrical, but separate architectures, whereby these differ only at the decoder level.
Let us first introduce the common backbone and then describe the specific decoder networks.

\noindent \textbf{GN Backbone.} Given a graph representation of a city (Figure \ref{fig:3-step-framework}), each node is assigned a feature representation of the state space $v_i$ (Section \ref{subsec:amod_rebalancing_mdp}) and a hidden state of a gated recurrent unit (GRU) $h_i$ \cite{ChoEtAl2014}.
First, we compute an embedding of raw node features through a feed-forward neural network $\psi_{\theta_e}$.
For each node, we generate an updated embedding $v_i^{'}$ and next hidden state $h_i^{'}$ through a GRU cell.
Finally, we aggregate information at a node level based on the topology of the graph (i.e., we extract sender $s_j$ and receiver $r_j$ information from the adjacency matrix).

\begin{algorithm}[t]
\caption{Temporal GN architecture}\label{alg:gn}
\begin{algorithmic}[1]
\Require Graph, $\mathcal{G} = (\{v_i\}, \{e_j, s_j, r_j\})$, $\mathcal{G}_h = (\{h_i\}, \{e_j^h, s_j, r_j\})$
\colorbox{pink}{\parbox{\linewidth}{
\For{each node $\{v_i\}, \{h_i\}$}
    \State compute node encoding $\psi_{\theta_e}(v_i)$ \hspace{12mm} Graph Encoder
    \State compute GRU recursion $(v_i^{'}, h_i^{'}) = f_n(\psi_{\theta_e}(v_i))$
\EndFor
}}
\colorbox{SkyBlue}{\parbox{\linewidth}{
\For{each node $\{v_i^{'}\}, \{h_i^{'}\}$}
    \State aggregate $h_i^{*}$ per receiver  $h_i^{*} = \sum_{j/r_j}h_i^{'}$
    \State aggregate $v_i^{*}$ per receiver  $v_i^{*} = \sum_{j/r_j}v_i^{'}$ \hspace{4.5mm} Graph Pooling
\EndFor
}}
\colorbox{LimeGreen}{\parbox{\linewidth}{
\If{actor}
    \State compute $\mathbf{a}_t \sim \text{Dirichlet}(\ba_t | \alpha), \alpha = \psi_{\theta_a}(\mathbf{v}^{*}, \mathbf{v})$ 
\EndIf
\If{critic} \hspace{42mm} Graph Decoder
    \State compute $v(\bs_t) = \psi_{\theta_c}(\mathbf{v}^{*}, \mathbf{v})$ 
\EndIf
}}
\Ensure Graph, $\mathcal{G}^{'} = (\{v_i^{*}\}, \{e_j, s_j, r_j\})$, $\mathcal{G}_h = (\{h_i^{*}\}, \{e_j^h, s_j, r_j\})$, $\{\ba_t, v(\bs_t)\}$
\end{algorithmic}
\end{algorithm}

\noindent \textbf{Decoder.} Given the updated graph representation, let us define two separate decoder architectures for the actor and critic.
As introduced in Section \ref{sec:control_of_single_city_amod_systems}, a rebalancing action is defined as the desired distribution of idle vehicles across all stations.
Thus, to obtain a valid probability density over actions, we define the output of our policy network to represent the concentration parameters $\mathbf{\alpha} \in \mathbb{R}_{+}^{N_v}$ of a Dirichlet distribution, such that $\ba_t \sim \text{Dir}(\ba_t | \mathbf{\alpha})$, and where the positivity of $\mathbf{\alpha}$ is ensured by a Softplus nonlinearity.
On the other hand, the critic is characterized by an additional \emph{global} sum-pooling performed on the updated graph representation.
In this way, the critic computes a single value function estimate for the entire network by aggregating information across all nodes in the graph.
For both architectures, we use skip-connection at the decoder level (i.e., by concatenating raw and updated node features, $[v^{*}, v_i]$).

\section{Experiments}
\label{sec:experiments}
In this section, we present simulation experiments\footnote{Code available at: \emph{\url{https://github.com/DanieleGammelli/graph-meta-rl-for-amod}}} using data from different cities around the world.
The goal of our experiments is to answer the following questions: (1) Can we learn a single AMoD controller capable of generalizing to unseen cities through meta-RL? (2) How do policies learned through meta-RL compare to traditional transfer learning strategies? (3) What are the adaptation capabilities of behavior policies learned through meta-RL?


\subsection{Benchmarks} 

All RL modules were implemented using PyTorch \cite{PaszkeGrossEtAl2019} and the IBM CPLEX solver \cite{ios_ILOG:1987} for the dispatching and minimal rebalancing cost problems.
For a complete description of the experimental details, please refer to Appendix \ref{app:A1}.
In our experiments, we compare the proposed Meta-RL framework with the following methods:

\noindent \textbf{Heuristics.} In this class of methods, we focus on measuring performance of simple, domain-knowledge-driven rebalancing heuristics.
    \begin{enumerate}[leftmargin=*]
        \item \emph{Random policy}: at each timestep, we sample the desired distribution from a Dirichlet prior with concentration parameter $\alpha = [1, 1, \ldots, 1]$. 
This benchmark provides a lower bound of performance within the three-step framework by choosing desired distributions of idle vehicles randomly.
        \item \emph{Equally distributed policy (ED)}: rebalancing actions are chosen to recover an equal distribution of vehicles across all areas in the transportation network.
    \end{enumerate}

\noindent \textbf{Learning-based.} Within this class of methods, we focus on measuring the generalization capabilities of common transfer learning strategies for RL agents.
For all methods, the A2C architecture is kept fixed (i.e., based on recurrent GNs), thus the difference lies exclusively on the learning strategy used at training time.
    \begin{enumerate}[leftmargin=*]
        \setcounter{enumi}{2}
        \item \emph{Zero-shot}: we train a single model on all meta-training tasks by iteratively sampling among the available cities. At test time, the model is deployed on unseen cities without any additional fine-tuning.
        \item \emph{Fine-tune}: as for the zero-shot case, we train a single model on all meta-training tasks by iteratively sampling among the available cities. At test time, we allow the model to have $n$ episodes - or equivalently, gradient steps - of interaction with the new environment, thus giving the model the chance to adapt to the novel environment through fine-tuning.
        \item \emph{Single-city}: this benchmark serves as an upper bound of performance within the RL class. Specifically, we assume to have unlimited access to the environment at both training \emph{and} test time, thus fully training a single AMoD controller from scratch, and until convergence, for every environment.
    \end{enumerate}

\begin{table*}[t]
\centering
\begin{tabular}{l c c | c c c c | c c}
 & & & & \multicolumn{2}{c}{GNN-RL} & & \\
 & Random & ED & Zero-shot & Fine-tune & Meta-RL & Single-city & MPC-Forecast & MPC-Oracle\\
\hline
 San Francisco  & 12.3 ($\pm$ 0.4) & 14.1 ($\pm$ 0.4) & 14.3 ($\pm$ 0.4) & 14.6 ($\pm$ 0.3) & \textbf{\highest{15.2 ($\pm$ 0.3)}} & 14.4 ($\pm$ 0.4) & 13.9 ($\pm$ 0.5) & 15.9 ($\pm$ 0.4)\\
 New York Brooklyn & 26.5 ($\pm$ 5.5) & 49.6 ($\pm$ 0.9) & 46.4 ($\pm$ 1.0) & 46.0 ($\pm$ 0.7) & \highest{56.2 ($\pm$ 0.7)} & 56.4 ($\pm$ 0.8) & \textbf{56.5 ($\pm$ 1.2)} & 57.2 ($\pm$ 0.8) \\
Shenzhen West & 48.6 ($\pm$ 1.7) & 58.4 ($\pm$ 1.3) & 60.3 ($\pm$ 1.1) & 60.9 ($\pm$ 0.8) & \highest{61.5 ($\pm$ 0.7)} & \textbf{63.5 ($\pm$ 0.9)} & 60.6 ($\pm$ 1.0) & 65.2 ($\pm$ 1.0)\\
\hline
\end{tabular}
\caption{Average reward (profit, thousands of dollars) on meta-test environments. \textbf{Black-bold} and \highest{red} highlight best performing (non-oracle) model and best performing \emph{adaptive} model, respectively. \textbf{\highest{Red-bold}} is used in case the two coincide.}
\label{tab:meta_learning_across_cities}
\end{table*}

\noindent \textbf{MPC-based.} Within this class of methods, we focus on measuring performance of MPC approaches that serve as state-of-art benchmarks for the rebalancing problem. 

\begin{enumerate}[leftmargin=*]
    \setcounter{enumi}{5}

    \item \emph{MPC-Oracle}: the MPC is based on a tri-level embedded optimization model that assumes perfect foresight information of future user requests and network conditions (e.g., travel times, prices, etc.).
In this formulation, the upper-level model determines the optimal desired vehicle distribution, and the lower-level models are the matching problem (Eq.~\ref{eq:matching}) and the minimum rebalancing-cost problem (Eq.~\ref{eq:reb}) that characterize the impact of the optimal desired vehicle distribution on the other modules within the three-step framework. Therefore, this approach serves as an \emph{oracle} that provides a performance upper bound for any algorithm within the three-step framework. Notice that MPC-Oracle does not scale well as the number of stations increases, since solving the embedded tri-level optimization model is NP-hard \cite{BenBlair1990}.
\item \emph{MPC-Forecast}: we relax the assumption of perfect foresight information in MPC-Oracle, substituting with a noisy and unbiased estimate of demand. 
This estimate takes the form of the rate of the underlying time-dependent Poisson process describing passenger arrival in the system.
This approach is a realistic control-based benchmark in the context of unknown system dynamics, but also exhibits poor scalability like MPC-Oracle.
\end{enumerate}

\subsection{Meta-learning Across Cities}
\label{subsec:exp_1}
In our first simulation experiment, we study system performance on meta-test environments.
Specifically, we train agents on taxi record data from the cities of Washington DC, Chicago, New York (Manhattan), Porto, Rome, Shenzhen (Baoan, East Downtown) and later simulate the deployment on the transportation systems for the cases of San Francisco, New York (Brooklyn), and Shenzhen (West Downtown).  Please refer to Appendix \ref{app:A1} for experimental details (e.g., data description and training specifics). 

Results in Table \ref{tab:meta_learning_across_cities} show that Meta-RL is able to achieve close-to-optimal performance on all three cities.
Specifically, the AMoD control policy learned through meta-RL is only 4.3\% (San Francisco), 1.8\% (New York) and 5.6\% (Shenzhen) away from oracle performance, which, as introduced in Section \ref{sec:experiments}, is assumed to have perfect information on the system dynamics.
Moreover, Meta-RL achieves by far the most robust performance when compared to other transfer strategies (i.e., Zero-shot and Fine-tune): up to a 18\% improvement in the case of New York, thus clearly highlighting the benefits of explicitly considering transfer and generalization in the design of training algorithms.
Results in Table \ref{tab:meta_learning_across_cities} also show how RL approaches (i.e., Meta-RL and Single-city) can achieve comparable, or better, performance with respect to MPC-Forecast, thus showcasing an advantage of learning-based control policies in the presence of model error: a characteristic of fundamental importance in real-world settings, whereby defining robust models of transportation demand is typically a challenging task.
Moreover, computationally, RL approaches based on graph neural networks exhibit computational complexity linear in the number of nodes and graph connectivity, opposed control-based approaches which scale super-linearly in the number of edges \cite{Brand2019}, thus allowing for fast computation of rebalancing policies through forward-propagation of the learned policy $\pi(\ba_t | \bs_t)$.
Lastly, Meta-RL achieves comparable performance to the Single-city approach.
However, a crucial aspect in the comparison between Meta-RL and Single-city is not the final performance in and of itself, rather, the performance of the two approaches as a function of iterations allowed with the environment at test time.
Specifically, if on one hand Single-city was trained until convergence on each meta-test environment individually (with $\approx 10,000$ episodes of interaction per city), Meta-RL achieves similar performance with only $n=10$ episodes, a $1000\times$ increase in data efficiency at test time.

\begin{figure}[t]
      \centering
     \includegraphics[width=0.95\columnwidth]{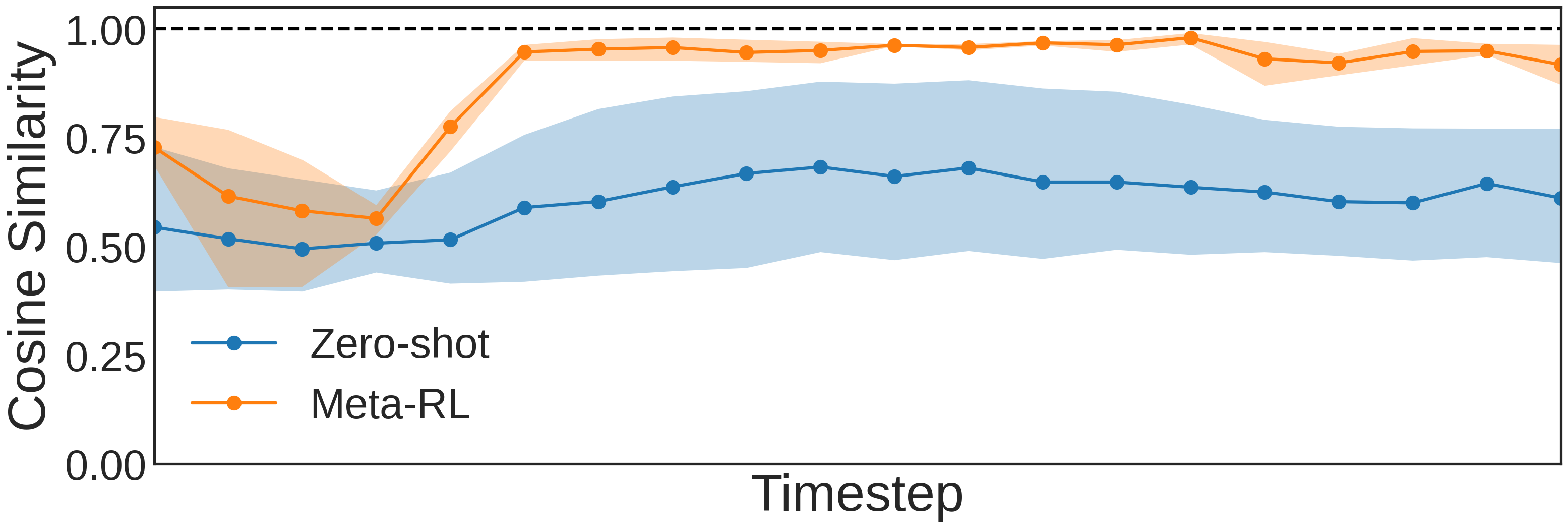}
      \caption{Measure of alignment between the agent's action and user demand. Confidence bounds are computed as an aggregation among the three meta-test environments.}
      \label{fig:cosine_similarity}
   \end{figure}

As a further analysis, let us try to better understand the learning algorithm acquired through meta-learning.
Concretely, we pose the following question: "what is the Meta-RL policy adapting to?".
To answer this question, we measure the alignment over time between the agent's actions (i.e., $\ba_t, t=1, \ldots, T$) and user demand in each station (i.e., $d_i^t = \sum_j d_{ij}^t, i \in \mathcal{V}$).
More formally, at each time $t$, we compute the cosine similarity between $\ba_t$ and $d_i^t$ (Figure \ref{fig:cosine_similarity}).
Results in Figure \ref{fig:cosine_similarity} highlight an interesting property of policies trained through meta-RL: as timesteps progress, actions strongly align with user demand ($\approx 0.9$ cosine similarity opposed to $\approx 0.6$ of Zero-shot approaches).
Given the absence of any predictive model of user demand within the state representation, Meta-RL showcases an interesting degree of \emph{implicit demand estimation}.
Specifically, having an implicit estimation of demand is a property that nicely aligns with intuition: in order to quickly adapt to arbitrary AMoD scenarios, learning to estimate demand from small task-related experience is a very advantageous skill to acquire.

\subsection{Online Adaptation to Disturbances}
\label{subsec:exp_2}
To further assess the generalization capabilities of Meta-RL, we also study the extent to which policies learned through meta-learning are able to adapt to system disturbances online.
Specifically, in this simulation experiment, we focus on the New York Brooklyn scenario and generate three common unexpected disturbances within any real-world mobility system: the presence of a peak in user demand given by a special event (Section \ref{subsubsec:exp_21}), a sudden increase in travel times in specific areas of the city because of traffic congestion (e.g., caused by accidents; Section \ref{subsubsec:exp_22}), and a change in pricing scheme (Section \ref{subsubsec:exp_23}). 
Within this set of experiments, the following remarks are made in order.
First, it is worth highlighting a fundamental difference between MPC-Oracle and all other approaches in presence of \emph{unexpected} disturbances.
Specifically, given the assumption that MPC-Oracle has access to perfect information on future states of the system (i.e. demand, travel times, prices, etc.) this approach has the possibility to \emph{plan} for the disturbance, thus serving as upper bound of adaptation. 
On the other hand, all other models either rely on a model-free approach by observing only the current state of the system (i.e. RL models), or on erroneous forecasts of future states (i.e. MPC-Forecast).
Because of this, all non-oracle approaches are forced to \emph{react} to the disturbance at test time.
Second, the Single-city model is not being re-trained under the disturbance, rather we use the parameter configuration used in Section \ref{subsec:exp_1} for the New York Brooklyn scenario.

\subsubsection{Adaptation to special events}
\label{subsubsec:exp_21}
In this experiment, we simulate the effects of a special event (e.g., concert, sporting event, etc.) in order to measure the robustness of Meta-RL to sudden changes in user demand distribution.
Specifically, we select the area of Forest Hills, Queens, and simulate the presence of a special event, such as a concert in the Forest Hills Stadium (Figure \ref{fig:policy_special_event}).
Results in Table \ref{tab:special_event}, show how Meta-RL is by far more robust than any other RL strategy (50.3\% improvement with respect to Single-city, and 22.1\% and 20.2\% from Zero-shot and Fine-tune, respectively), achieving better system performance when compared to MPC-Forecast controls under model error (4.6\% improvement).
In particular, these results highlight how the common approach in literature of training of RL agents on single-city scenarios might lead to learning AMoD control policies very sensitive to changes in distribution.

Qualitatively, Figure \ref{fig:policy_special_event} shows how Meta-RL exhibits a behavior that nicely aligns with intuition, thus reacting to the unexpected increase in demand in order to maximize its performance. 


\subsubsection{Changes in Pricing Scheme}
\label{subsubsec:exp_22}
To further assess how well Meta-RL can adapt online to changes in distribution, we now consider the scenario where the operator modifies its pricing scheme (e.g., for regulation compliance or higher profits). Notice that we assume pricing and fleet management are separate modules, as commonly used in existing systems.  
We simulate a change in prices for all outbound and inbound flow from the three of the most active areas in Brooklyn, i.e., the surroundings of John F. Kennedy Airport, Queens Plaza and LaGuardia Airport.
Enforced by the change in pricing scheme, any control policy will need to radically change its behavior in order to fulfill the goal of maximizing operator profit.
Concretely, we simulate an average $\times5$ reduction in prices for all passenger flow within the area of LaGuardia Airport, and a $\times2$ increase in prices for the areas of JKF Airport and Queens Plaza.

Table \ref{tab:price} and Figure \ref{fig:policy_price_change} show how Meta-RL is able to outperform all other non-oracle models, with only a 4.7\% drop in performance when compared to MPC-Oracle.
Specifically, Meta-RL achieves a 35.1\%, 35.7\% and 11.6\% improvement when compared Zero-shot, Fine-tune and Single-city approaches.
By observing the aggregated performance of average served demand and rebalancing cost, it is interesting to observe how Meta-RL incurs a higher rebalancing cost (e.g. almost double compared to Zero-shot and Fine-tune) while still achieving similar levels of served customers, thus showcasing a more sophisticated selection of trips in order to maximize profit.


\subsubsection{Congestion}
\label{subsubsec:exp_23}
We now investigate the ability of Meta-RL to adapt to sudden increases in travel time (e.g., due to accidents or unexpected weather condition).
We simulate a situation of high congestion for all inbound traffic within the area of JFK Airport, which in our experiments is the most active and profitable area in Brooklyn.
In particular, we simulate an average $\times4$ increase in inbound travel times.
By doing so, a naive policy could run the risk of having many vehicles stuck in traffic due to excessive rebalancing, thus reducing vehicle utilization and hindering profitability.

Results in Table \ref{tab:congestion} and Figure \ref{fig:policy_congestion} show how Meta-RL achieves better system performance when compared to all other learning and heuristic-based approaches.
In this specific scenario, Meta-RL incurs a 9\% drop in performance compared to MPC-Forecast.
Specifically, in the case of travel times, the use of recursive planning together with the ability to observe current congestion levels enables MPC-Forecast to achieve satisfactory system performance despite an erroneous estimate of future travel times.
Finally, this scenario once again highlights the limitations of the traditional single-city training paradigm: the resulting policies perform very well under the training distribution, but rapidly degrade in the presence of unexpected disturbances.


\begin{table}[t]
\centering
\begin{tabular}{l c c c}
    & Reward & Served Demand & Reb. Cost\\
    \hline 
    Random & 20.7 ($\pm$ 3.9) & 1,513 & 23.1 \\ [0.1ex]
    ED & 43.8 ($\pm$ 1.3) & 2,387 & 8.6 \\[0.1ex]
    \hline
    Zero-shot & 38.8 ($\pm$ 1.0) & 1,642 & 7.9\\ [0.1ex]
    Fine-tune & 39.7 ($\pm$ 1.6) & 1,626 & 8.1 \\[0.1ex]
    Meta-RL & \textbf{\highest{49.8 ($\pm$ 1.4)}} & 2,558 & 15.6  \\[0.1ex]
    Single-city & 24.7 ($\pm$ 5.0) & 1,707 & 11.0  \\[0.1ex]
    \hline
    MPC-Forecast & 47.5 ($\pm$ 2.4) & 2,975 & 16.2 \\[0.1ex]
    MPC-Oracle & 57.5 ($\pm$ 0.8) & 3,111 & 7.7 \\[0.1ex]
    \hline
    \end{tabular}%
\caption{Special event scenario in New York Brooklyn. \textit{Reward} is profit, and \textit{Reb. Cost} is rebalancing cost.}
\label{tab:special_event}%
\end{table}

\begin{table}[t]
\centering
\begin{tabular}{l c c c}
    & Reward & Served Demand & Reb. Cost\\
    \hline 
    Random & 107.4 ($\pm$ 13.8) & 2,835 & 26.3 \\ [0.1ex]
    ED & 139.2 ($\pm$ 3.3) & 3,570 & 11.3 \\[0.1ex]
    \hline
    Zero-shot & 129.8 ($\pm$ 3,707) & 3,367 & 9.4 \\ [0.1ex]
    Fine-tune & 128.7 ($\pm$ 3,139) & 3,262 & 8.4 \\[0.1ex]
    Meta-RL & \textbf{\highest{200.2 ($\pm$ 4.5)}} & 3,835 & 14.8  \\[0.1ex]
    Single-city & 177.0 ($\pm$ 8.1) & 4,072 & 15.0  \\[0.1ex]
    \hline
    MPC-Forecast & 199.0 ($\pm$ 5.3) & 4,040 & 15.9 \\[0.1ex]
    MPC-Oracle & 210.2 ($\pm$ 5.2) & 3,989 & 14.3 \\[0.1ex]
    \hline
    \end{tabular}%
\caption{Price change scenario in New York Brooklyn}
\label{tab:price}%
\end{table}

\begin{table}[t]
\centering
\begin{tabular}{l c c c}
    & Reward & Served Demand & Reb. Cost \\
    \hline 
    Random & 3.0 ($\pm$ 3.3) & 1,687 & 24.9 \\ [0.1ex]
    ED & 22.5 ($\pm$ 1.5) & 2,092 & 16.4 \\[0.1ex]
    \hline
    Zero-shot & 22.3 ($\pm$ 0.7) & 2,000 & 13.6\\ [0.1ex]
    Fine-tune & 22.2 ($\pm$ 1.1) & 2,000 & 13.6 \\[0.1ex]
    Meta-RL & \highest{27.5 ($\pm$ 1.2)} & 2,315 & 16.5  \\[0.1ex]
    Single-city & 3.3 ($\pm$ 2.3) & 2,780 & 34.1  \\[0.1ex]
    \hline
    MPC-Forecast & \textbf{30.2 ($\pm$ 1.1)} & 2,152 & 10.2 \\[0.1ex]
    MPC-Oracle & 36.2 ($\pm$ 928) & 2,192 & 4.2 \\[0.1ex]
    \hline
    \end{tabular}%
\caption{Congestion scenario in New York Brooklyn}
\label{tab:congestion}%
\end{table}

\subsection{Sensitivity to Number of Training Tasks}

\begin{figure}[t]
    \centering
    \includegraphics[width=0.95\columnwidth]{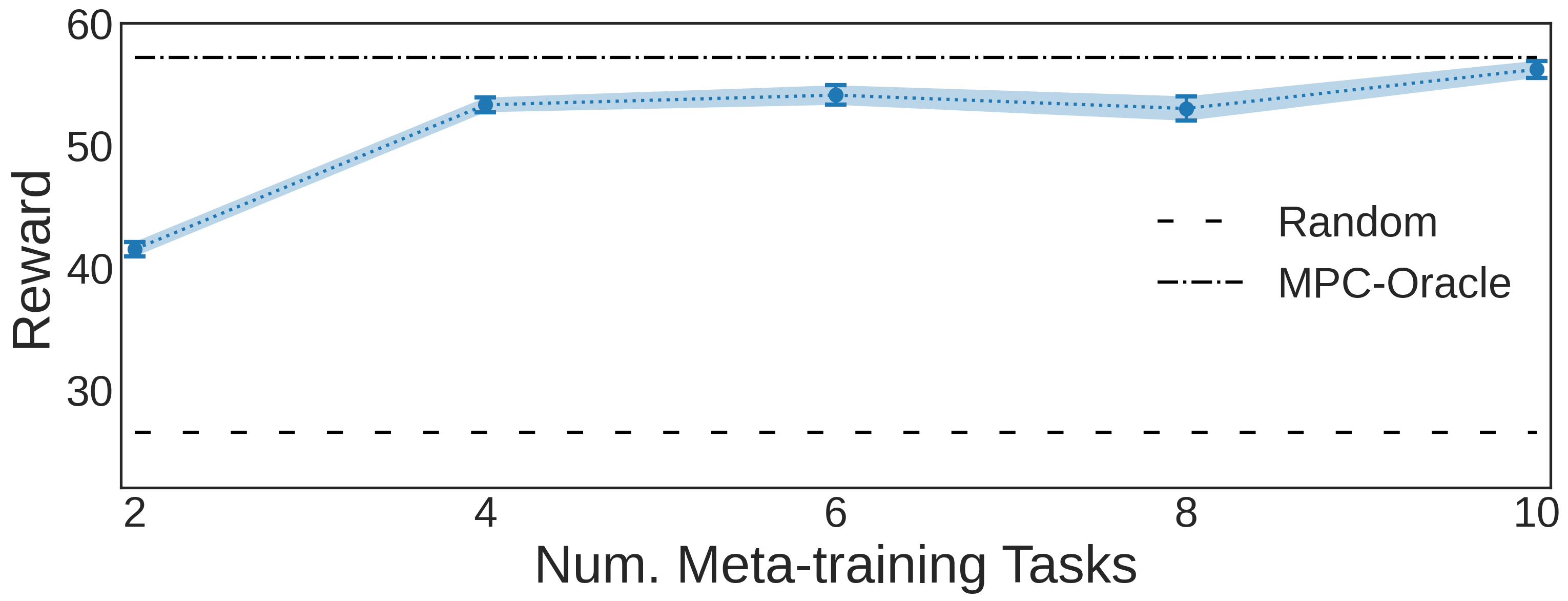}
    \caption{Reward with respect to the number of meta-training tasks for the New York Brooklyn scenario.}
    \label{fig:sensitivity}
\end{figure}

In Figure \ref{fig:sensitivity}, we analyze the effect of the number of tasks on the performance of behavior policies learned through meta-RL.
Results clearly show how system performance improves as the number of tasks (i.e., cities) increases.
In particular, analogous to the training process for supervised learning frameworks, the meta-learning process treats tasks as data samples and the performance of meta-training algorithms relies on having a diverse set of meta-training tasks.
Recently, a number of regularization strategies for meta-learning have been proposed \cite{JamalEtAl2018}, and we believe these results show evidence that it is possible to consider strategies to avoid meta-overfitting in the context of AMoD systems, thus representing an interesting and fruitful direction for future work.

\section{Conclusion}
\label{sec:conclusion}
This work addresses the problem of synthesizing control strategies for AMoD systems from small amount of experience.
We argue that current research efforts should move away from the traditional single-city treatment of the AMoD control problem and focus on devising learning paradigms capable of guaranteeing robust system performance across a diverse set of geographies and, more generally, changes in distribution.
In particular, we believe the ability to leverage previous experience from e.g., cities in which the system is already operational, will be a key enabling factor for the deployment of real-world systems.
To do so, we cast controlling multi-city AMoD systems as a meta-RL problem and devise a deep RL agent based on recurrent graph neural networks for the task.
Crucially, we show how explicitly considering transfer and generalization in the design of the neural architectures and training strategies can enable RL agents to recover highly adaptive behavior policies.

This research opens several promising future directions. First, it is interesting to investigate ways of pre-training RL agents exclusively through offline data \cite{LevineEtAl2020}, thus removing all interactions with the environment until a sufficient level of performance is guaranteed.
Second, we would like to explore architectures to address increased complexity and stochasticity in the system dynamics (e.g., endogenous congestion, elastic demand). 
Through their ability to learn about the environment, RL approaches represent a promising direction for learning to control challenging environments such as the ones described by complex human-robot interactions.

\label{subsec:sensitivity}

\begin{acks}
The authors would like to thank M. Tsao, D. Jalota, and A. Faust for insightful comments and M. Zallio for help with the graphics. This research was partially supported by the National Science Foundation
(NSF) under the CPS program (\#1837135) and NTT Docomo Inc. K. Yang would like to acknowledge the Swiss  National  Science  Foundation (SNSF) Postdoc Mobility Fellowship (P400P2\_199332). This article
solely reflects the opinions and conclusions of its authors and not NSF, Docomo, SNSF, or any other entity.
\end{acks}


\bibliographystyle{unsrt} 
\bibliography{main,ASL_papers}

\newpage
\appendix
\onecolumn
\section*{APPENDIX}
\setcounter{section}{1}

\subsection{Experimental Details}
\label{app:A1}
\textbf{Models.} \hspace{4mm}
We train each model using stochastic gradent ascent to optimize the expected sum of discounted rewards $\mathbb{E}_{\bs_t, \ba_t \sim \pi} \left[\sum_t \gamma^t r(\bs_t, \ba_t) \right]$, with discount factor $\gamma = 0.97$ and an episode length $T=20$.
We use the Adam optimizer with a fixed learning rate of $0.0003$ for both actor and critic architectures.
We train Single-city, Zero-shot and Fine-tune for 10,000 episodes and Meta-RL for 1,000 trials of $n=10$ episodes.
At test time, we allow Fine-tune to have $10$ episodes of interaction with the environment.
In our proposed GN architecture, the node encoder $\psi_{\theta_e}(\cdot)$ is a single-layer feed-forward neural network with $256$ hidden states followed by a ReLU non-linearity.
The GRU cell is also characterized by a 256-dimensional hidden state.
We use a sum aggregation function to aggregate information for all nodes in the graph.
The updated node representation is then mapped to either Dirichlet concentration parameters (i.e., with dimensionality equal to the number of nodes in the respective graph) or value function estimate (i.e., 1-dimensional) through a single-layer feed-forward neural network.

\noindent \textbf{Data.} The case studies are based on the taxi record data collected in San Francisco \cite{PiorkowskiSarafijanovicEtAl2009}, Washington DC \cite{WashingtonDC2019}, New York City \cite{TaxiLimousineCommission2013}, Chicago \cite{Chicago2013}, Rome \cite{BraccialeBonolaEtAl2014}, Porto \cite{MoreiraGamaEtAl2013}, and Shenzhen\cite{ZhangZhaoEtAl2015}. These cities across continents are chosen to represent different network topologies and demand patterns. Two larger cities, New York City and Shenzhen, are further divided into four regions such that the inter-regional demand is minimized. In each scenario, the road network is segmented into stations by clustering junctions such that the travel time within each station is upper-bounded by a given error tolerance (e.g., a time step). The trip record data are converted to demand, travel times, and trip prices between stations. Here, the demand (Poisson rate of customer arrival) is aggregated from the trip record data every 15 minutes. Descriptive statistics of the scenarios used are presented in Table \ref{tab:statistics}.


 \begin{table*}[t]
\centering
\begin{tabular}{l c c c c c c}
    & Date and time & No. Nodes	& Max. Trip Time  & Avg. Trip Time 	& Avg. Demand  &  No. Vehicles \\
    & & & (min) & (min) & (req./hr) & 
    \\\hline  
    Porto & 2013-11-26 8:00-9:00 & 13 & 15 & 6 & 1952 & 240\\
    NYC Manhattan North & 2013-03-08 19:00-20:00 & 12 & 18 & 6 & 4555 & 800 \\
    NYC Manhattan Center & 2013-03-08 19:00-20:00& 12 & 25 & 7 & 7686 & 1500 \\
    NYC Manhattan South & 2013-03-08 19:00-20:00& 14 & 33 & 8 & 7968 & 1500 \\
    NYC Brooklyn & 2013-03-08 19:00-20:00 & 14 & 68 & 18 & 3162 & 1500 \\
    Chicago & 2014-03-15 19:00-20:00 & 11 & 50 & 13 & 11446 & 2729 \\
    Washington DC & 2019-03-12 19:00-20:00 & 18 & 68 & 14 & 1000 & 1097 \\
    San Francisco & 2008-06-06 8:00-9:00 & 10 & 55 & 11 & 1380 & 374 \\
    Rome & 2014-02-04 8:00-9:00 & 13 & 69 & 18 & 177 & 79 \\
    Shenzhen Baoan & 2013-10-22 8:00-9:00 & 15 & 66 & 22 & 2372 & 918 \\
    Shenzhen North  & 2013-10-22 8:00-9:00& 23 & 66 & 19 & 2581 & 867 \\
    Shenzhen Downtown West  & 2013-10-22 8:00-9:00 & 17 & 66 & 15 & 4637 & 1777 \\
    Shenzhen Downtown East  & 2013-10-22 8:00-9:00 & 12 & 65 & 15 & 3422 & 1141 \\
 \hline 
    \end{tabular}%
\caption{Scenario statistics}
\label{tab:statistics}%
\end{table*}

\subsection{Adaptation to Disturbances: Qualitative Analysis}
\begin{figure*}[ht]
      \centering
     \includegraphics[width=\textwidth]{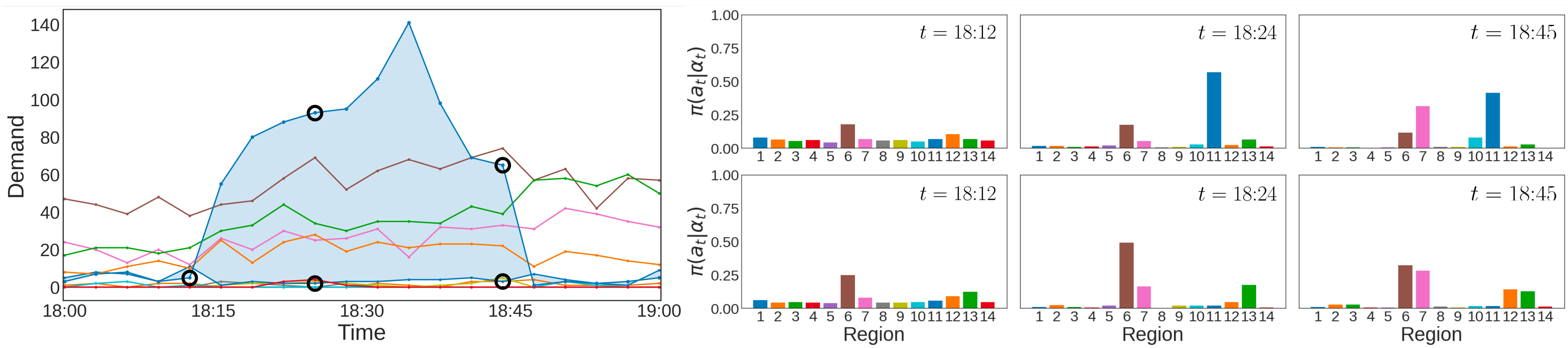}
      \caption{Policy visualization for the Special Event scenario. (Left) Time-series of user demand. The blue shaded region shows the increase in transportation requests compared to the average demand in Forest Hills. For all remaining areas, we assume overall decreased traffic given the presence of the special event. (Right) Visualization of the Meta-RL policy's output (i.e., desired distribution of idle vehicles) in three specific time-steps (black circles highlight the corresponding demand observations). Results show how Meta-RL is able to substantially adapt its behavior in presence of the special event (top row), compared to a standard use-case, i.e., without special event (bottom row).}
      \label{fig:policy_special_event}
   \end{figure*}

\begin{figure*}[ht]
      \centering
     \includegraphics[width=\textwidth]{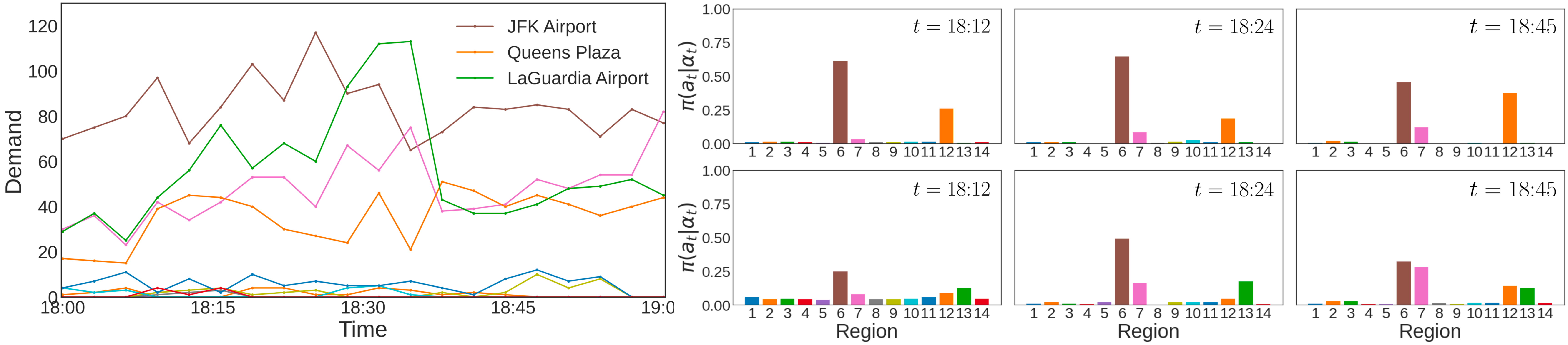}
      \caption{Policy visualization for the Price Change scenario. (Left) Time-series of user demand. Here we focus on three of the most active areas: JFK and LaGuardia Airports (approximately $\times2$ increase and $\times5$ decrease in prices, respectively) and Queens Plaza ($\times2$ increase in prices). (Right) Visualization of the Meta-RL policy's output (i.e., desired distribution of idle vehicles) in three specific time-steps. Results show how Meta-RL is able to substantially adapt its behavior in presence of the price change (top row), compared to a standard use-case, i.e., without price change (bottom row), by favoring more profitable demand.}
      \label{fig:policy_price_change}
   \end{figure*}

\begin{figure*}[ht]
      \centering
     \includegraphics[width=\textwidth]{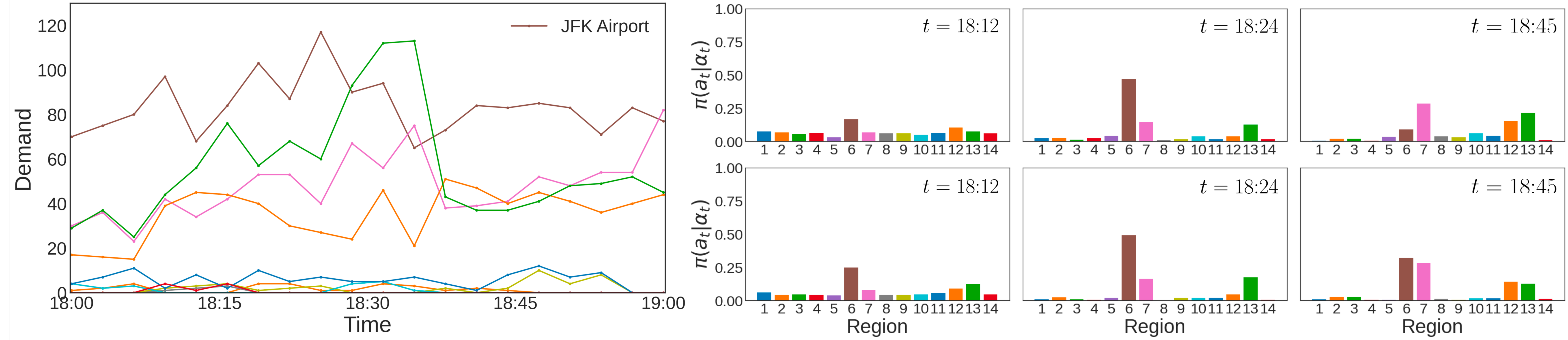}
      \caption{Policy visualization for the Congestion scenario. (Left) Time-series of user demand. Here we focus on the area of JFK Airport (approximately $\times4$ increase in travel times for all inbound and outbout flow). (Right) Visualization of the Meta-RL policy's output (i.e., desired distribution of idle vehicles) in three specific time-steps. Results show how Meta-RL is able to substantially adapt its behavior in presence of the congestion (top row), compared to a standard use-case, i.e., without congestion (bottom row), by avoiding to over-saturate JFK with the risk of having many vehicles stuck in traffic.}
      \label{fig:policy_congestion}
   \end{figure*}

In Figures \ref{fig:policy_special_event}, \ref{fig:policy_price_change}, and
\ref{fig:policy_congestion} we report a qualitative visualization of the learned AMoD control policies and their capacity to react to unexpected changes in distribution.

\end{document}